\documentclass[aps,prl,final,twocolumn,superscriptaddress,%
floatfix,preprintnumbers]{revtex4}
\usepackage{dcolumn} 
\usepackage{amssymb}
\usepackage{amsbsy}
\usepackage[tbtags]{amsmath}

\usepackage{graphicx}
\usepackage{floatflt}
\usepackage{hyperref}  

\usepackage{overpic}

\usepackage{xcolor}
\usepackage{ulem}
\usepackage{mciteplus}
\usepackage{wrapfig}

\newcommand{\bc}{\begin{center}}
\newcommand{\ec}{\end{center}}
\def\t0{t\text{$=$}0}
\def\ix0{\xi\text{$=$}0}

\newcommand{\bea}{\vspace{-0mm}\begin{eqnarray}}
\newcommand{\eea}{\end{eqnarray}}

\newcommand{\fm}{\operatorname{fm}}

\newcommand{\units}[1]{\ensuremath{\,\mathrm{#1}}}

\newcommand{\elll}{l}

\newcommand{\prp}{\perp}

\newcommand{\ren}{{\text{ren}}}

\newcommand{\tcdot}{{\cdot}}
\newcommand{\ldotP}{{\elll\!\cdot\!P}}

\newcommand{\myRe}{\ensuremath{\mathrm{Re}}}
\newcommand{\myIm}{\ensuremath{\mathrm{Im}}}

\newcommand{\vect}[1]{\ensuremath{\boldsymbol{#1}}}
\newcommand{\vprp}[1]{\vect{#1}_\prp}

\hypersetup{
  debug=false,                                    
  pdfpagemode={UseOutline},                       
  pdftitle={Intrinsic quark transverse momentum in the nucleon from lattice QCD},
  pdfauthor={Hagler Musch Negele Schafer},
  colorlinks=True,
  citecolor = {blue},
  filecolor = {blue},
  urlcolor = {blue},
  linkcolor = {blue}
}

\begin{document}  

\title{Intrinsic quark transverse momentum in the nucleon from lattice QCD}

\preprint{TUM/T39-09-08, MIT-CTP 4056}

\author{Ph.~H\"agler}
  \affiliation{Institut f\"ur Theoretische Physik T39,
   Physik-Department der TU M\"unchen, 85747 Garching, Germany}
   \email{phaegler@ph.tum.de}
\author{B.U.~Musch}
  \affiliation{Institut f\"ur Theoretische Physik T39,
   Physik-Department der TU M\"unchen, 85747 Garching, Germany}
\author{J.W.~Negele}
  \affiliation{Center for Theoretical Physics, Massachusetts Institute of Technology, Cambridge, Massachusetts 02139, USA}
\author{A.~Sch\"afer}
  \affiliation{Institut f\"ur Theoretische Physik, Universit\"at
  Regensburg, 93040 Regensburg, Germany}

\date{\today}

\begin{abstract} 
A better understanding of transverse momentum ($\vprp{k}$-) 
dependent quark distributions in a hadron is needed to interpret several experimentally observed 
large angular asymmetries and to clarify the fundamental role of gauge links in non-abelian gauge theories. 
Based on manifestly non-local gauge invariant quark operators we introduce process-independent 
$\vprp{k}$-distributions and study their properties in lattice QCD.
We find that the longitudinal and transverse momentum dependence approximately factorizes, 
in contrast to the behavior of generalized parton distributions. 
The resulting quark $\vprp{k}$-probability densities for the nucleon show characteristic dipole deformations
due to correlations between intrinsic $\vprp{k}$ and the quark or nucleon spin.
Our lattice calculations are based on $N_f{=}2{+}1$ mixed action propagators of the LHP collaboration.
\end{abstract}

\maketitle

{\em Introduction.}---
Already 30 years ago, it has been noted that intrinsic transverse momentum, $\vprp{k}$, of partons
gives rise to azimuthal asymmetries in unpolarized semi-inclusive deep inelastic scattering (SIDIS), 
for example $e^-{+}p{\rightarrow} e^-{+}\pi{+}X$, nowadays known as the Cahn effect \cite{Cahn:1978se}.
Since then, significant progress has been made in understanding intrinsic $\vprp{k}$ effects
and their relation to the eikonal phases that quark fields acquire in hadron scattering processes
due to initial and final state interactions \cite{Brodsky:2002cx,*Belitsky:2002sm}.
The eikonal phases, given by gauge links (Wilson lines), turn out to be \emph{process-dependent} and lead to, e.g., the
Sivers and Collins asymmetries \cite{Sivers:1989cc,Collins:1992kk}  
in polarized SIDIS, which
have attracted a lot of attention and were already observed in experiments at HERMES, COMPASS and Jefferson Lab 
\cite{HERMES:2009ti,*COMPASS:2008dn,*Avakian:2005ps}.
Theoretically, these 
\begin{wrapfigure}{r}{0.2\textwidth}
\bc
\vspace{-4mm}
\includegraphics[width=0.2\textwidth]{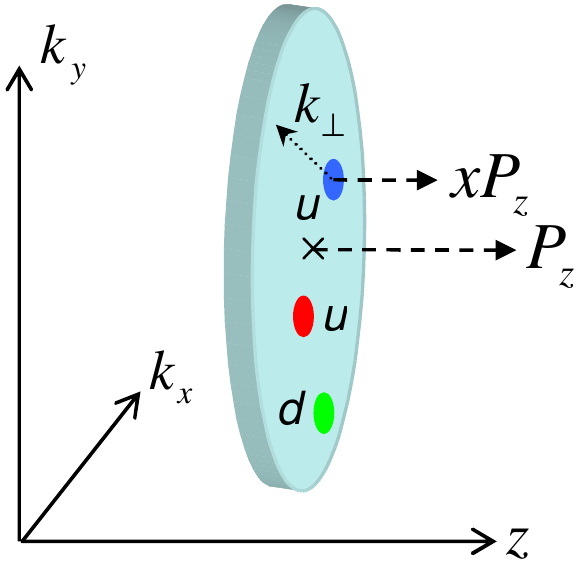}
\caption{\label{FigIll} Illustration of the transverse momentum distribution of quarks in the proton.}
\ec
\vspace{-4mm}
\end{wrapfigure}
\noindent can be described in the framework of QCD factorization using 
transverse momentum dependent parton distribution functions (tmdPDFs) 
\cite{Collins:1992kk,Ji:2004wu}, an approach that goes beyond the usual collinear approximation and operator
product expansion 
involving (moments of) PDFs.
In addition to their phenomenological importance, tmdPDFs provide essential information about the internal structure of hadrons
in the form of probability densities in the transverse momentum plane, 
$\rho(x,\vprp{k})$, as illustrated in Fig.~\ref{FigIll} \cite{comment1}, 
where $x$ is the longitudinal 
momentum fraction 
carried by the quark.

In this work, we introduce process-independent $\vprp{k}$-distributions and calculate these in lattice QCD.
We illustrate our results by presenting $\vprp{k}$-densities of quarks
in the nucleon, 
with a focus on possible correlations between $\vprp{k}$ and the transverse quark and nucleon spins,
resulting in deformations from a spherically symmetric distribution.
It is interesting to compare this approach with generalized parton distributions (GPDs) in impact parameter ($\vprp{b}$-) 
space \cite{Burkardt:2000za},
which allows one to study
the \emph{spatial} distribution of partons in hadrons in form of probability densities $\rho(x,\vprp{b})$ \cite{Diehl:2005jf}.
Lattice QCD studies of the latter revealed characteristic non-spherical shapes of the pion and the nucleon 
in the case of transversely polarized quarks \cite{Gockeler:2006zu,Brommel:2007xd}.
We stress, however, that tmdPDFs and GPDs provide fundamentally different and complementary insight into hadron structure, 
since they are not related by Fourier transformation  and  $\vprp{k}$ and $\vprp{b}$ are not conjugate variables.

To introduce the different tmdPDFs, we first define the
momentum-space correlators
$\Phi_{\Gamma}{=}\Phi_{\Gamma}(x,\vprp{k};P,S)$,
\bea
	\Phi_{\Gamma}\!&=&
	\!\int\! d(\bar n\tcdot k) \!\int\! \frac{d^4l}{2(2\pi)^4} e^{-ik \cdot l} \widetilde\Phi_{\Gamma}(l;P,S)\nonumber\\
	&=&\!\int\! d(\bar n\tcdot k) \!\int\! \frac{d^4l}{2(2\pi)^4} e^{-ik \cdot l}\langle P,S| \bar q(l) \Gamma \mathcal{U} q(0)| P,S\rangle \;.
	\label{tmdPDFs1}
\eea
with nucleon states $|P,S\rangle$ depending on momentum and spin, 
and where the Wilson line $\mathcal{U}{=}\mathcal{U}_{\mathcal{C}(l,0)}$, defined by a path ordered exponential,
ensures gauge invariance of the non-local quark operator $\bar q(l) \ldots q(0)$.
For the vector (unpolarized), $\Gamma_V^\mu{=}\gamma^\mu$, axial-vector (polarized), $\Gamma_A^\mu{=}\gamma^\mu\gamma_5$, 
and tensor (quark helicity flip), $\Gamma^{\mu \nu}_T{=}i\sigma^{\mu \nu}\gamma_5$, cases, 
the correlators in Eq.~\ref{tmdPDFs1} can be parametrized by the 
twist-2 tmdPDFs \cite{Mulders:1995dh,*Boer:1997nt}:
\bea
	 n_\mu \Phi^\mu_V &=& f_1 + \vect{S}_i \vect{\epsilon}_{\perp ij} \vect{k}_j \frac{1}{m_N}\, f_{1T}^\perp \nonumber\\
	 n_\mu \Phi^\mu_A &=& 
	\Lambda g_{1}	+ \frac{\vprp{k} \cdot \vprp{S}}{m_N}\ g_{1T}\nonumber\\
	 n_\mu \Phi^{\mu j}_T &=&     
   -  \vect{S}_j h_1^{\phantom{\perp\!\!}}
   -  \frac{\vect{\epsilon}_{\perp ji} \vect{k}_i}{m_N}\, h_{1}^\perp   
\nonumber \\
  &-&  \frac{\Lambda \vect{k}_j }{m_N}\, h_{1L}^\perp - 
  \frac{(2 \vect{k}_j \vect{k}_i - \vprp{k}^2 \delta_{ji}) \vect{S}_i}{2m_N^2}\, h_{1T}^\perp \;,
	\label{tmdPDFs2}
\eea
where the distributions $f,g,h$ depend on $x$ and $\vprp{k}$ and $\Lambda$ is the nucleon helicity.
The light-cone vectors $n$ and $\bar n$ in Eqs.~\ref{tmdPDFs1},\ref{tmdPDFs2} are chosen such that
$n\tcdot\bar n{=}1$, $P{=}P^+ \bar n + m_N^2/(2P^+)n$, and the 
contraction with $n_\mu$ in Eq.~\ref{tmdPDFs2} projects on leading twist for large $P^+$.
Formally \cite{comment1}, three of the tmdPDFs are directly related to the well-known unpolarized, $f_1(x)$, polarized, $g_1(x)$, 
and transversity, $h_1(x)$, PDFs by $[f_1,g_1,h_1](x){=}\int d^2\vprp{k} [f_1,g_1,h_1](x,\vprp{k})$.
Further (approximate) relations between tmdPDFs and GPDs have been established 
based on a dynamical mechanism \cite{Burkardt:2002ks}, 
by analogy comparing $\vprp{k}$- and 
$\vprp{b}$-densities \cite{Diehl:2005jf},
and in the framework of quark models \cite{Meissner:2007rx}.
Such relations exist in particular for the Sivers, $f_{1T}^\perp$, and Boer-Mulders, $h_{1}^\perp$, functions, 
which are \emph{naively} time-reversal odd. 
Remarkably, the tmdPDFs $g_{1T}$ and $h_{1L}^\perp$ cannot be directly related to any of the GPDs for reasons of time reversal symmetry,
so that their appearance may be seen as \emph{genuine} sign of intrinsic $\vprp{k}$ of quarks.
In particular, they cannot be generated dynamically from coordinate ($\vprp{b}$-) space 
densities by final state interactions.
We emphasize that the definition of the correlator in Eq.~\ref{tmdPDFs1} and therefore the tmdPDFs depends on the gauge link path $\mathcal{C}(l,0)$.
Apart from the precise form of $\mathcal{C}(l,0)$, a further issue
in the QCD factorization of SIDIS and Drell-Yan processes
is the appearance of so-called 
rapidity divergences, which require regularization or subtraction and may be seen in
analogy to the Bloch-Nordsieck theorem (see \cite{Collins:2008ht} and references therein).
In the case of SIDIS, the path is generically given by $\mathcal{C}(l,0){=}[l,l+\infty n,\infty n,0]$, which is not invariant
under time reversal, and hence the appearance of the Sivers and Boer-Mulders functions.
There exist two natural choices for process-independent $\vprp{k}$-distributions,
employing either a direct Wilson line $\mathcal{C}(l,0){=}[l,0]$,
or an average over all possible gauge links between $0$ and $l$. 
We chose the first possibility for this pioneering study.
Our results should not be directly compared to phenomenological studies of 
SIDIS \cite{Anselmino:2005nn} or the Drell-Yan process.
In particular, the Sivers and Boer-Mulders functions vanish for direct Wilson lines.
To describe the corresponding 
asymmetries, we will have to use specific gauge link structures,
which is the subject of ongoing studies \cite{MuschThesis2,Musch:2009xyz}.
Lattice QCD gives access to hadron matrix elements 
$\widetilde\Phi_{\Gamma}(l;P,S)$ (cf. Eq.~\ref{tmdPDFs1}), 
which can be parametrized by complex-valued invariant amplitudes $\widetilde A_i(l^2,l\tcdot P)$.
The choice of a straight Wilson line, together with hermiticity, parity and time-reversal symmetry leaves
us with 8 independent amplitudes:
\bea
  \widetilde\Phi^\mu_{V}& =& 4  P^\mu \widetilde{A}_2
		+ 4im_N^2  l^\mu \widetilde{A}_3 \;,\nonumber\\
	\widetilde\Phi^\mu_{A}	& = &
		- 4 m_N  S^\mu \widetilde{A}_6
		- 4im_N P^\mu l \tcdot S \widetilde{A}_7 
		+ 4m_N^3  l^\mu l \tcdot S \widetilde{A}_8\nonumber\\
	\widetilde\Phi^{\mu\nu}_{T}	& = &
		4 S^{[\mu} P^{\nu]} \widetilde A_{9m} + 4im_N^2 S^{[\mu} l^{\nu]} \widetilde A_{10}\nonumber\\
	&-& 2 m_N^2 \left[ 2 l\tcdot S\; l^{[\mu} P^{\nu]}  {-} l^2  S^{[\mu} P^{\nu]}  \right] \widetilde A_{11}	\,,
	\label{amplitudes1}
\eea
where $[\mu \nu]{=}\mu\nu-\nu\mu$.
Detailed comparison with Eqs.~\ref{tmdPDFs2} reveals that the tmdPDFs are given by certain Fourier-integrals of the amplitudes, where the Fourier conjugate variables are $\ldotP$ and $x$, and $\vprp{l}$ and $\vprp{k}$.
Such relations exist between 
$f_{1}{\leftrightarrow} \widetilde A_{2}$,
$g_{1}{\leftrightarrow} \widetilde A_{6,7}$, 
$h_1{\leftrightarrow} \widetilde A_{9m}$,
$g_{1T}{\leftrightarrow}\widetilde A_{7}$,
$h^\perp_{1L}{\leftrightarrow}\widetilde A_{10,11}$,
and
$h^\perp_{1T}{\leftrightarrow}\widetilde A_{11}$, while $\widetilde A_{3,8}$ give only contributions to
higher-twist tmdPDFs.

\begin{figure}[t]
	\centering 
	\setlength{\unitlength}{0.01\linewidth}
	\begin{picture}(100,100)
		\put(2,2){\includegraphics[width=0.98\linewidth,clip=true,trim=27 32 47 48]{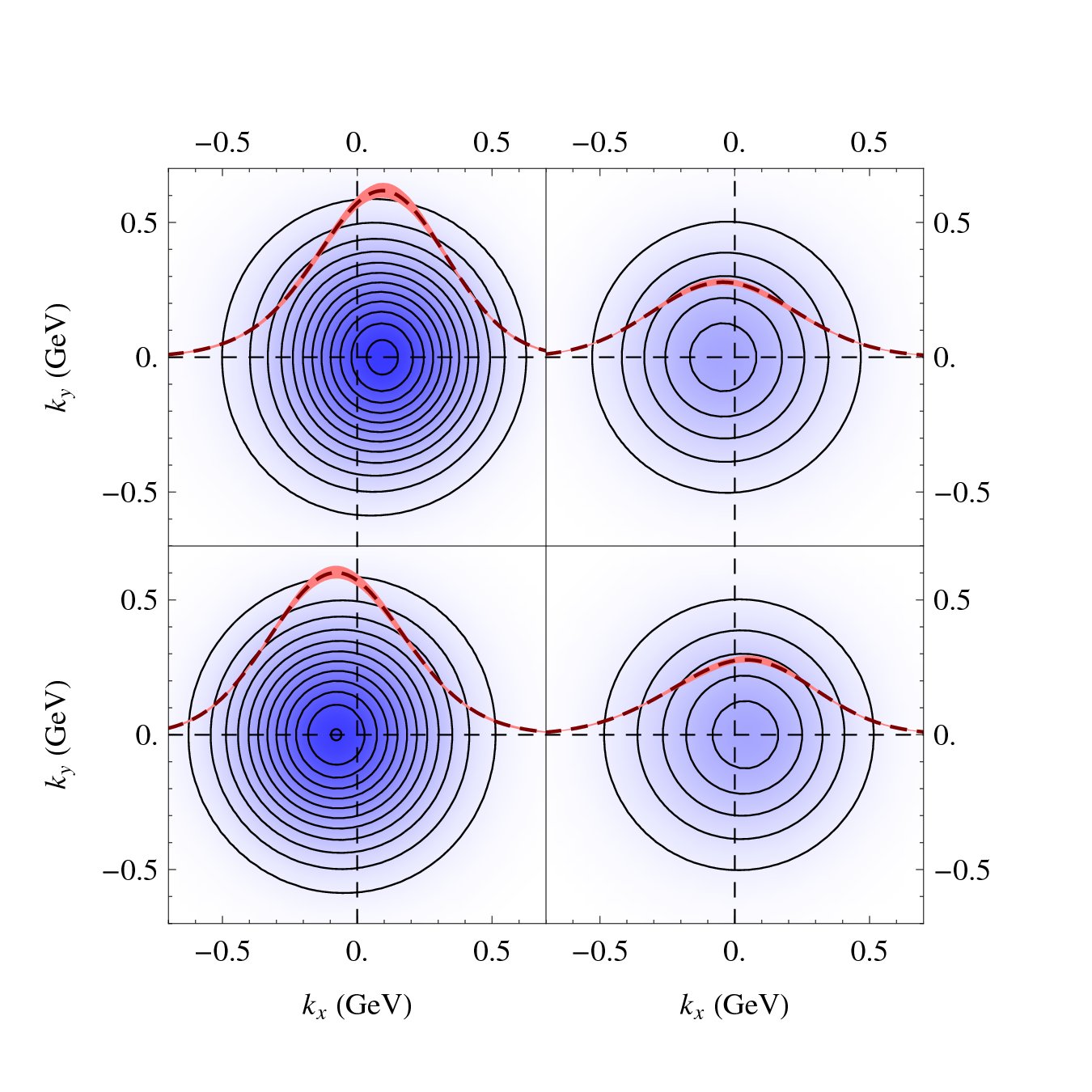}}
		\put(0,74){$\vect{k}_y$}
		\put(0,30){$\vect{k}_y$}
		\put(31,0){$\vect{k}_x$}
		\put(75,0){$\vect{k}_x$}
		\put(13,92){(a)}
		\put(13,54){\includegraphics[height=1cm]{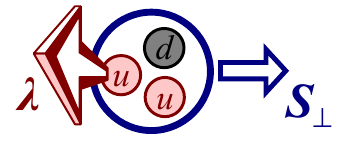}}
		\put(57,92){(b)}
		\put(57,54){\includegraphics[height=1.167cm]{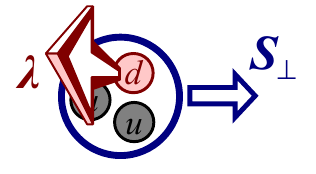}}
		\put(13,47){(c)}
		\put(13,10){\includegraphics[height=1cm]{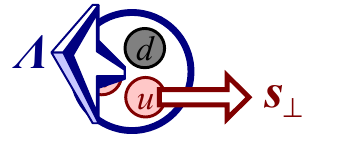}}
		\put(57,47){(d)}
		\put(57,10){\includegraphics[height=1cm]{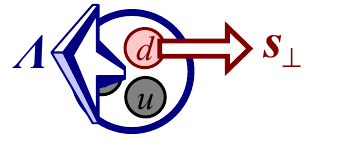}}
		\end{picture}
	\caption{Quark densities in the $\vprp{k}$-plane, for $m_\pi {\approx} 500\units{MeV}$. 
	(a) $\rho_L$ for $u$-quarks and $\lambda=1$, $\vprp{S}=(1,0)$, (b) the same 
	for $d$-quarks, (c) $\rho_T$ for $u$-quarks and $\Lambda=1$, $\vprp{s}=(1,0)$, 
	(d) the same for $d$-quarks. 
	The error bands show the density profile at $\vect{k}_y=0$ as a function of $\vect{k}_x$ (scale not shown). }
	\label{densities1}
\end{figure}
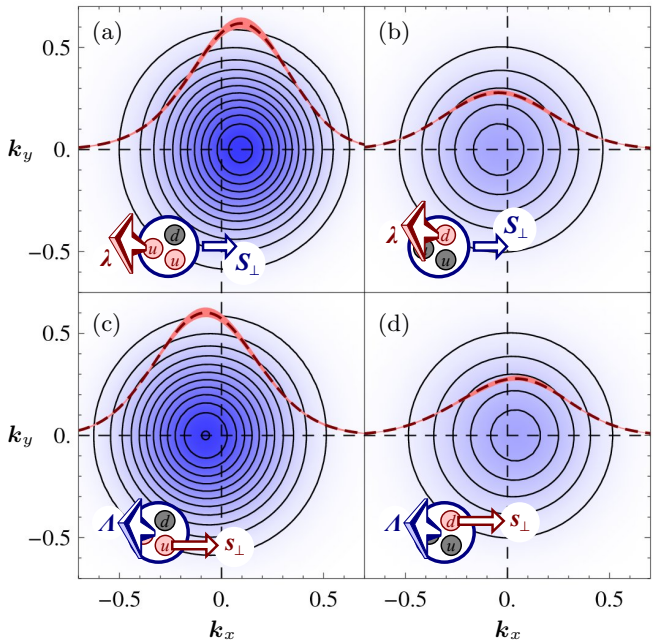

{\em Lattice QCD results.}---
In our lattice study, we 
use a discretized, Euclidean version of the non-local operator 
$\mathcal{O}_\Gamma(\elll) {=} \bar q(\elll)\, \Gamma\, \mathcal{U}_{[\elll,0]}\, q(0)$. 
Limiting ourselves to purely 
spatial quark separations, $\elll^0 {=} 0$, the operator resides on a plane of equal Euclidean time. 
A product of connected link variables is used to represent the straight Wilson line $\mathcal{U}_{[\elll,0]}$. 
For oblique angles, we approximate a straight line by a step-like (zig-zag) path.
According to the continuum analysis of 
Ref.~\cite{Dorn:1986dt}, the renormalized operator 
can be written as $O_\Gamma^\ren(\elll) {=} \mathcal{Z}^{-1} \exp(-\delta m\, \sqrt{-\elll^2})\ O_\Gamma(\elll)$, involving two 
renormalization constants, $\mathcal{Z}^{-1}$ and $\delta m$, the latter being associated with a potential power divergence 
of the Wilson line. 
The renormalization of Wilson lines in a cut-off regularized theory,
e.g. on the lattice, and the appearance of power divergences of the form $l/a$ 
has been studied extensively in the context of heavy quark effective theory, see, e.g., 
\cite{Maiani:1991az,*Martinelli:1995vj}. 
We utilize the fact that $\delta m$ also appears 
in the renormalization of the static quark potential: $V^\ren(r) {=} V(r) + 2 \delta m$. 
At large $r$, the static quark potential is well approximated by the string potential $V^\ren_\text{string}(r) {=} \sigma r - \pi/12 r + C^\ren$. 
To fix $\delta m$ we follow Refs. \cite{Cheng:2007jq,*Bazavov:2009zn} and match $V^\ren(r)$ to the string potential at $r=1.5r_0{\approx}0.7\fm$, 
setting $C^\ren{=}0$.
Note that $\vprp{k}$-distributions are sensitive to the renormalization condition implied by the choice of $C^\ren$.
In our numerical computations, we employ MILC gauge configurations \cite{Ber01} based on the AsqTad improved staggered quark action with 
two light and one heavier (strange) quark flavors.
For our exploratory calculations,
we have chosen an ensemble at a pion mass of ${\approx}500\units{MeV}$ and a lattice spacing of $a{=}0.124\units{fm}$.
From an analysis of $V(r)$ from Wilson loops on the HYP smeared gauge configurations, 
we obtain $a \delta m{=} -0.155(5)_\text{stat}$ \cite{MuschThesis2,Musch:2009xyz}.
To determine the amplitudes $\tilde A_i$, we take advantage of the methods and techniques  
used in Ref. \cite{Hagler:2007xi} for the calculation of GPDs. 
As in Ref. \cite{Hagler:2007xi}, we only include contributions from quark line connected diagrams. 
A major innovation in our analysis is that the lattice correlation (three-point) functions are
evaluated for manifestly non-local operators $\mathcal{O}_\Gamma(\elll)$, hence requiring 
a parametrization according to Eq.~\ref{amplitudes1}.
They are numerically evaluated using 
the CHROMA library \cite{Edwards:2004sx} for domain wall quark propagators and 
sequential propagators previously calculated by the LHP collaboration, 
with the valence quark mass tuned to match the staggered sea quarks \cite{Hagler:2007xi}. 
The sequential propagators 
are available for two lattice nucleon momenta $\vect{P}{=}(0,0,0)$ and $\vect{P}{=}(-1,0,0)$, 
the latter corresponding to ${\approx}500\units{MeV}$ in physical units. 
We consistently use the nucleon mass of $m_N {=} 1.291(23)\units{GeV}$ obtained for this ensemble in the extraction of the
amplitudes and tmdPDFs from Eqns.~(\ref{tmdPDFs2},\ref{amplitudes1}).
All errors quoted are statistical
(see Refs.~\cite{MuschThesis2,Musch:2009xyz} for a more comprehensive treatment of uncertainties).
Setting $l^0{=}0$, we
are restricted to the region given by $\elll^2<0$ and $|\elll \tcdot P| \leq \sqrt{-\elll^2} |\vect{P}|$.
Since only finite hadron momenta $\vect{P}$ can be accessed on the lattice,
this precludes us from evaluating the full $x$- and $\vprp{k}$-dependence directly. 
An important result, demonstrated in detail by our lattice data in \cite{MuschThesis2,Musch:2009xyz}
in the kinematic region accessible to us, 
is factorization of the $x$ and $\vprp{k}$ ($\ldotP$ and $l^2$) dependence to a good approximation.
Factorization is a standard assumption for which we have now provided quantitative support, 
and for this reason we will focus 
in the following on the $l^2$ dependence and present results for the lowest $x$-moments corresponding to $l\cdot P$ = 0, e.g.,
\begin{figure}[t]
\vspace{-2mm}
	\setlength{\unitlength}{0.01\linewidth}
	\begin{picture}(100,86)
		\put(0,3){\includegraphics[width=\linewidth,clip=true,trim=0 22 0 0]{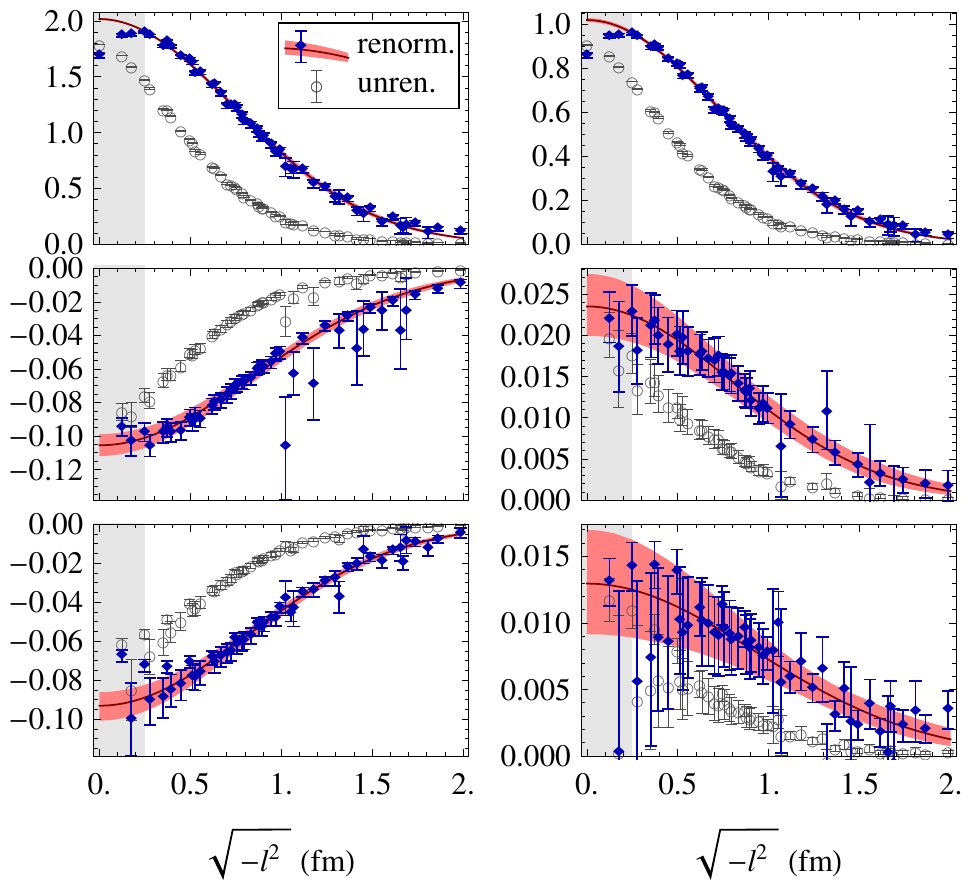}}
		\put(23,0){$\sqrt{-\elll^2}$}
		\put(73,0){$\sqrt{-\elll^2}$}
		\put(11,62){\begin{minipage}[b]{15ex}\flushleft(up)\\$\myRe\ \tilde A_2$\par\end{minipage}}
		\put(28.5,36){$\myRe\ \tilde A_7$ (up)}
		\put(27,10){$\myRe\ \tilde A_{10}$ (up)}
		\put(74,79){$\myRe\ \tilde A_2$ (down)}
		\put(74,53){$\myRe\ \tilde A_7$ (down)}
		\put(73,27){$\myRe\ \tilde A_{10}$ (down)}
		\end{picture}
	\caption{Real parts of the amplitudes $\tilde A_i(\elll^2,\elll \tcdot P{=}0)$, for $m_\pi {\approx} 500\units{MeV}$. 
	The solid lines and error bands
	 are Gaussian fits to the renormalized lattice data above $\sqrt{-\elll^2}{=}0.25\units{fm}$.}
  \label{amplitudes}
\end{figure}
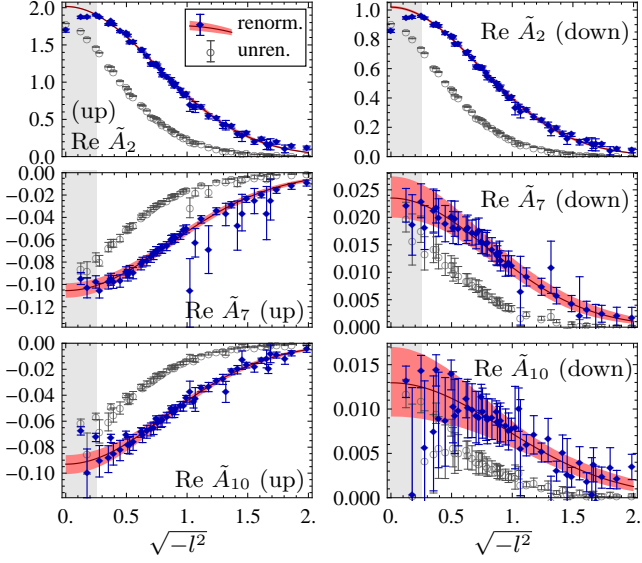
%
%
\begin{equation*}
	f_1^{(0)}(\vprp{k}) \equiv \int_{-1}^{1}\hspace{-1.5ex} dx\, f_1(x,\vprp{k}^2) = \int d^2 \vprp{l}\, e^{i\vprp{l}\tcdot \vprp{k}} 2 \widetilde A_2(-\vprp{l}^2,0).
\end{equation*}
We note that the $x$-integrated distributions of type $f^{(0)}$ and $h^{(0)}$ 
correspond to the difference, while distributions of type $g^{(0)}$ correspond to the sum
of quark- and antiquark distributions, respectively.
Hermitian conjugation shows that $\myIm\widetilde A_i(l^2,l\tcdot P{=}0){=}0$.
Representative results for $\myRe \widetilde A_{2,7,10}(-\vprp{l}^2,0)$ are displayed in Fig.~\ref{amplitudes}. 
Similar results not shown in this work have been obtained for the amplitudes $\widetilde A_{6,9m,11}$.
We have parametrized our renormalized lattice data 
using a Gaussian ansatz
$2 \widetilde A_i^{u,d}(-\vprp{\elll}^2,0) {=} c_i^{u,d} \exp(- \vprp{\elll}^2/\sigma_i^{u,d})$.
Corresponding fits to the lattice data are represented by the error bands in Fig.~\ref{amplitudes}.
To avoid contamination by lattice cut-off effects at small $l\lessapprox a$, we exclude data points for $\sqrt{-l^2}\leq0.25\fm$ 
from the fit.
The Gaussians smoothly bridge the excluded region at small $\elll^2$. 
Thus we do not resolve 
potential divergences of the amplitudes $A_i(\elll^2,0)$ in the continuum limit at 
$\elll^2{\rightarrow}0$. 
In that sense, the Gaussian parametrization effectively acts as a  
regularization prescription 
at short distances $\elll^2$, which translates into a smooth cutoff of the resulting $\vprp{k}$-distributions at large $\vprp{k}^2$. 
Within this context, we 
define $\vprp{k}^2$-moments 
\begin{equation*}
	[f,g,h]^{(0,n)} =  \int d^2\vprp{k}\, (\vprp{k}^2/2 m_N^2)^n\, [f,g,h]^{(0)}(\vprp{k}^2).
\end{equation*}
Since $f^{(0)}_1(\vprp{k}^2)$ is the unpolarized density of quarks minus antiquarks, we can 
fix $\mathcal{Z}$ by demanding that the total number of 
$u$- minus $d$-quarks in the proton is $f^{(0,0)}_{1,u-d}{=}c_2^{u-d}=1$, 
giving $\mathcal{Z}{=}1.056(14)$. 
\begin{table}
\vspace{-1mm}
	\renewcommand{\arraystretch}{1.4}
	\begin{tabular}{|c||ll|l|}
	\hline
	& \multicolumn{2}{c|}{$c$} & \multicolumn{1}{c|}{$2/\sigma\ (\mathrm{GeV})$} \\ \hline \hline
	$\tilde A_2^u$    & \phantom{-}2.0159(86)   & $=f_{1,u}^{(0,0)}$     & 0.3741(72) \\
	$\tilde A_2^d$    & \phantom{-}1.0192(90)   & $=f_{1,d}^{(0,0)}$     & 0.3839(78) \\
	$\tilde A_6^u$    &           -0.920(35)    & $=-g_{1,u}^{(0,0)}$   & 0.311(11) \\
	$\tilde A_6^d$    & \phantom{-}0.291(19)    & $=-g_{1,d}^{(0,0)}$   & 0.363(18) \\
	$\tilde A_{9m}^u$ & \phantom{-}0.931(29)    & $=h_{1,u}^{(0,0)}$     & 0.3184(90) \\
	$\tilde A_{9m}^d$ &           -0.254(16)    & $=h_{1,d}^{(0,0)}$     & 0.327(15) \\
	$\tilde A_7^u$    &           -0.1055(66)   & $=-g_{1T,u}^{(0,1)}$   & 0.328(14) \\
	$\tilde A_7^d$    & \phantom{-}0.0235(38)   & $=-g_{1T,d}^{(0,1)}$   & 0.346(36) \\
	$\tilde A_{10}^u$ &           -0.0931(73)   & $=h_{1L,u}^{\prp(0,1)}$   & 0.340(14) \\ 
	$\tilde A_{10}^d$ & \phantom{-}0.0130(40)   & $=h_{1L,d}^{\prp(0,1)}$   & 0.301(48) \\ 
	\hline
	\end{tabular}
	\renewcommand{\arraystretch}{1.0}
	\caption{
		Parameters obtained from a fit of the amplitudes $\widetilde A_i(-\vprp{\elll}^2,0)$ to Gaussian functions,
	  for $m_\pi {\approx} 500\units{MeV}$. 
		\label{fitparam}
		}
\end{table}
The Gaussian fit parameters are listed in Table \ref{fitparam}. 
Note that the width $\sigma$ is very sensitive to our renormalization condition $C^\ren = 0$.
For $\widetilde A_{2,6,9m}$, the inverse width $2/\sigma{=}\langle \vprp{k}^2 \rangle^{1/2}$ has an interpretation as the  
root mean square transverse momentum of the respective distribution $f_1^{(0)}$, $g_{1}^{(0)}$, $h_{1}^{(0)}$.
The fact that $g_{1,u-d}^{(0,0)}{=}1.209(36)$ comes out close to the physical $g_A{=}1.2695(29)$ 
and agrees within errors with the direct calculation in \cite{Edwards:2005ym} represents a remarkable 
validation of our approach.
Interestingly, we find that $g_{1T}^{(0,1)}{\approx} -h_{1L}^{\prp(0,1)}$, which supports corresponding results
obtained in quark model calculations \cite{Jakob:1997wg,*Pasquini:2008ax}.
The $\vprp{k}$-densities of quarks in the nucleon for 
longitudinally ($L$) and transversely polarized quarks ($T$) can be obtained from
$\rho_L{=}\rho(x,\vprp{k};\lambda,\vprp{S}) {=} n\tcdot(\Phi_{V}+\lambda\Phi_{A})/2$
and $\rho_T{=}\rho(x,\vprp{k};\Lambda,\vprp{s}) {=} (n\tcdot\Phi_{V}-n_\mu s_j\Phi^{\mu j}_{T})/2$,
respectively, and 
appear in form of a multipole-expansion \cite{Diehl:2005jf}
\bea
	\rho_{L} &=&
	\frac{1}{2} \bigg( f^{}_1 + \lambda\Lambda g_1 
	  + \bigg[ \frac{\vect{S}_j \vect{\epsilon}_{ji} \vect{k}_i}{m_N}\, f_{1T}^\perp \bigg]
    + \lambda\frac{\vprp{k} \cdot \vprp{S}}{m_N} g_{1T} \bigg)\nonumber\\
	\rho_{T} &=&
	\frac{1}{2} \bigg( f_1 
	 + \vprp{s}\cdot \vprp{S} h_1
   + \bigg[ \frac{\vect{s}_j\vect{\epsilon}_{ji} \vect{k}_i}{m_N}\, h_{1}^\perp\bigg]  \nonumber \\
   &+&  \Lambda\frac{\vprp{k} \cdot \vprp{s}}{m_N} h_{1L}^\perp
   + \frac{\vect{s}_j(2 \vect{k}_j \vect{k}_i - \vprp{k}^2 \delta_{ji}) \vect{S}_i}{2m_N^2}\, h_{1T}^\perp \bigg)
	\,,
	\label{rho1}
\eea
with monopole terms $\propto f_1,g_1,h_1$, dipole structures $\propto f_{1T},g_{1T},h_{1}^\perp,h_{1L}^\perp$, 
and a quadrupole term $\propto h_{1T}^\perp$.
The terms in square brackets proportional to the T-odd Sivers 
and Boer-Mulders functions are absent for straight Wilson lines.
We stress that the dipole-correlations of the types $\propto\lambda \vprp{k} \tcdot \vprp{S} g_{1T}$
and $\propto\Lambda \vprp{k} \tcdot \vprp{s} h_{1L}^\perp$ in Eqs.~\ref{rho1} are a characteristic feature of
intrinsic transverse momentum, and that analogous terms in the case of spatial (impact parameter) densities
are forbidden by time reversal symmetry \cite{Diehl:2005jf}.
Based on our results for the $\vprp{k}$-distributions, we show
in the upper part of Fig.~\ref{densities1} the lowest $x$-moment of the $\vprp{k}$-density $\rho_{L}$ of 
longitudinally polarized $u$- and $d$-quarks in a transversely polarized nucleon, with $\lambda{=}+1$ and 
$\vprp{S}{=}(1,0)$.
The densities feature significant
dipole deformations along the direction of the nucleon spin  
$\vprp{S}$, due to non-vanishing average transverse momentum shifts $\langle \vect{k}_x \rangle {=} m_N g_{1T}^{(0,1)}/f_{1}^{(0,0)}$.
We find sizeable shifts of $\langle \vect{k}_x \rangle {=} 67(5)\units{MeV}$ for $u$-, and 
$\langle \vect{k}_x \rangle {=} -30(5)\units{MeV}$ for $d$-quarks. 
An analogous observation is made for 
transversely polarized quarks in a longitudinally polarized nucleon, 
as displayed in the lower part of Fig.~\ref{densities1} for $\Lambda{=}+1$ and $\vprp{s}{=}(1,0)$. 
Here we find shifts of similar magnitude but opposite sign: 
$\langle \vect{k}_x \rangle {=} m_N h_{1L}^{\perp(0,1)}/f_{1}^{(0,0)} {=} -60(5)\units{MeV}$ 
for $u$-, and $\langle \vect{k}_x \rangle {=} 16(5)\units{MeV}$ for $d$-quarks.
We note that these intrinsic $\vprp{k}$-shifts are orthogonal 
to the dipole deformations $\propto\vect{s}_j\vect{\epsilon}_{ji} \vect{b}_i$ 
of densities in impact parameter space observed in \cite{Gockeler:2006zu}.
%
%

{\em Conclusions.}---
We studied tmdPDFs of the nucleon for the first time using lattice QCD.
We found that the $x$ and the $\vprp{k}$ dependence approximately factorizes and that
quarks in the nucleon carry significant intrinsic transverse momentum.
The $\vprp{k}$-densities for polarized quarks in a polarized nucleon
show characteristic dipole deformations along the transverse spin vectors.
This work represents a first milestone in the non-perturbative calculation of $\vprp{k}$-distributions.
Ongoing studies including non-straight Wilson lines promise in particular to give access to the highly interesting T-odd Sivers effect. 
\begin{acknowledgments}
The authors acknowledge support by the Emmy-Noether program and the cluster of excellence ``Origin and Structure of the Universe'' of the DFG 
(Ph.H. and B.M.), SFB/TRR-55 (A.S.) and the US Department of Energy grant DE-FG02-94ER40818 (J.N.). 
\end{acknowledgments}


\bibliography{TMDs}

\end{document}